\documentclass[
reprint,
amsmath,
superscriptaddress,
amssymb,
aps,
pra,
showkeys
]{revtex4-1}

\usepackage{graphicx}
\usepackage{bm}
\usepackage{hyperref}
\usepackage{braket}
\usepackage{afterpage}
\usepackage[english]{babel}
\usepackage{natbib}
\newcommand{\him}[1]{{\ensuremath{\rm #1}}}

\begin{document}
\preprint{APS/123-QED}
\title{Multimode thermal states with multiphoton subtraction: study of the photons number distribution in the selected subsystem}
\author{K. G. Katamadze}\email{kgk@quantum.msu.ru}
\affiliation{Quantum Technology Centre, Faculty of Physics, M. V. Lomonosov Moscow State University,119991, Moscow, Russia}
\affiliation{Valiev Institute of Physics and Technology, Russian Academy of Sciences,117218, Moscow, Russia}
\affiliation{National Research Nuclear University MEPhI,115409, Moscow, Moscow, Russia}
\author{G. V. Avosopiants}
\affiliation{Quantum Technology Centre, Faculty of Physics, M. V. Lomonosov Moscow State University,119991, Moscow, Russia}
\affiliation{National Research University of Electronic Technology MIET, 124498, Moscow, Russia}
\author{N. A. Bogdanova}
\affiliation{Valiev Institute of Physics and Technology, Russian Academy of Sciences,117218, Moscow, Russia}
\affiliation{National Research University of Electronic Technology MIET, 124498, Moscow, Russia}
\author{Yu. I. Bogdanov}
\affiliation{Valiev Institute of Physics and Technology, Russian Academy of Sciences,117218, Moscow, Russia}
\affiliation{National Research Nuclear University MEPhI,115409, Moscow, Moscow, Russia}
\affiliation{National Research University of Electronic Technology MIET, 124498, Moscow, Russia}
\author{S. P. Kulik}
\affiliation{Quantum Technology Centre, Faculty of Physics, M. V. Lomonosov Moscow State University,119991, Moscow, Russia}

\begin{abstract}
	Thermal states of light are widely used in quantum optics due to their correlation properties. As is well known, their correlation properties and the photon number distribution as a whole are strongly dependent on the mode number selected by the detection scheme. The same changes can be caused by photon subtraction. Therefore, we describe the general case of the multimode thermal state after a multiphoton subtraction, when the photon number statistics is registered by the detector selecting a part of the initial modes. We present an analytical form of the obtained photon number distribution and its general properties and check them in the experiment.
\end{abstract}
\maketitle
\section{Introduction}
	Historically, thermal states of light lay in the basis of quantum optics. In 1900 Max Plank while studying the black-body spectrum came to the concept of photons and in 1956 Robert Hanbury Brown and Richard Q. Twiss showed that photon number correlations really exist and can be utilized to measure the apparent angular size of stars \cite{HanburyBrown1956}. However, thermal states of light are classical i. e. they can be “described with customary visualization by considering a light beam as a set of waves” \cite{Klyshko1996}. Thus, in recent years scientists use them to probe some quantum phenomena in order to understand if these effects are really quantum, and if any non-classical properties of light give any benefits, or not. Among them: ghost imaging \cite{Strekalov1995,Gatti2004,Ferri2005,Valencia2005}, Hong-Ou-Mandel interference \cite{Hong1987,Liu2013}, Schmidt modes correlation \cite{Bobrov2013}, super resolution, based on multiphoton interference \cite{Classen2016}, etc. Lately thermal states utilization in effects based on the photon annihilation in several modes, like quantum vampire effect \cite{Fedorov2015,Katamadze2018,Katamadze2019}, Photonic Maxwell’s demon \cite{Vidrighin2016}, quantum thermal engine \cite{Hlousek2017} etc. became very attractive, so the general theory of multiphoton subtracted multimode thermal states has grown in relevance.
 Since photon subtraction significantly modifies the photon number distribution, this effect can be also useful for tailoring speckle intensity statistics \cite{Bromberg2014, Bender2018}. Moreover, the theory of photon-subtracted thermal state can be enhanced to non-degenerate squeezed vacuum state, because each mode of such state has thermal statistics, which can be tuned by means of photon subtraction \cite{Iskhakov2016}.

	Generally, the significance and relevance of thermal states is based on their correlation properties. Single-mode thermal state of light can be described by the density operator, which has a well-known diagonal form in the Fock basis \cite{Scully2001}:
	\begin{equation}\label{eq:rho_ts}
		\hat{\rho}_{TS}=\sum\limits_{n=0}^{\infty}P_{BE}(n|\mu_0)\ket{n}\!\!\bra{n},
	\end{equation}
	where $ P_{BE}(n|\mu_0)=\frac{\mu_0^n}{(1+\mu_0)^{n+1}} $  is a Bose-Einstein distribution with the mean photon number per mode $ \mu_0 $. Thus, thermal field is a typical representative of super-Poissonian light with a second-order correlation function $ g^{(2)}=2 $.
	
	Photon subtraction applied to the single-mode thermal state leads to correlation decrease. It has been theoretically shown \cite{Agarwal1992} and experimentally verified \cite{Allevi2010,Zhai2013,Bogdanov2017} that the photon number distribution of $ k $-photon subtracted thermal states can be described by a negative binomial or a compound Poisson distribution \cite{Bogdanov2003}
	\begin{equation}\label{eq:Pcp}
		P_{cP}(n|\mu_0,a)=\frac{\Gamma(a+n)}{\Gamma(a)}\frac{\mu_0^{n}}{n!}\left(\frac{1}{1+\mu_0}\right)^{n+a}, 	
	\end{equation}
	where $ a=k+1 $ is the coherence parameter, $ \mu=a\mu_0 $ is the mean photon number. The correlation function of the considered distribution $ g^{(2)}=1+\frac{1}{a} $. Therefore, the density matrix of the $ k $-photon subtracted thermal state equals:
	\begin{equation}\label{eq:rho_kts}
		\hat{\rho}_{kTS}=\sum\limits_{n=0}^{\infty}P_{cP}(n|\mu_0,k+1)\ket{n}\!\!\bra{n}.
	\end{equation}

	It is interesting, that the total photon number $ N $ of an $ M $-mode thermal state is subject to exactly the same compound Poisson distribution $ P_{cP}(N|\mu_0,a=M) $  \cite{Mandel1995} and has the same correlation properties.

\begin{figure}[t]
		\center{\includegraphics[width=0.8\columnwidth]{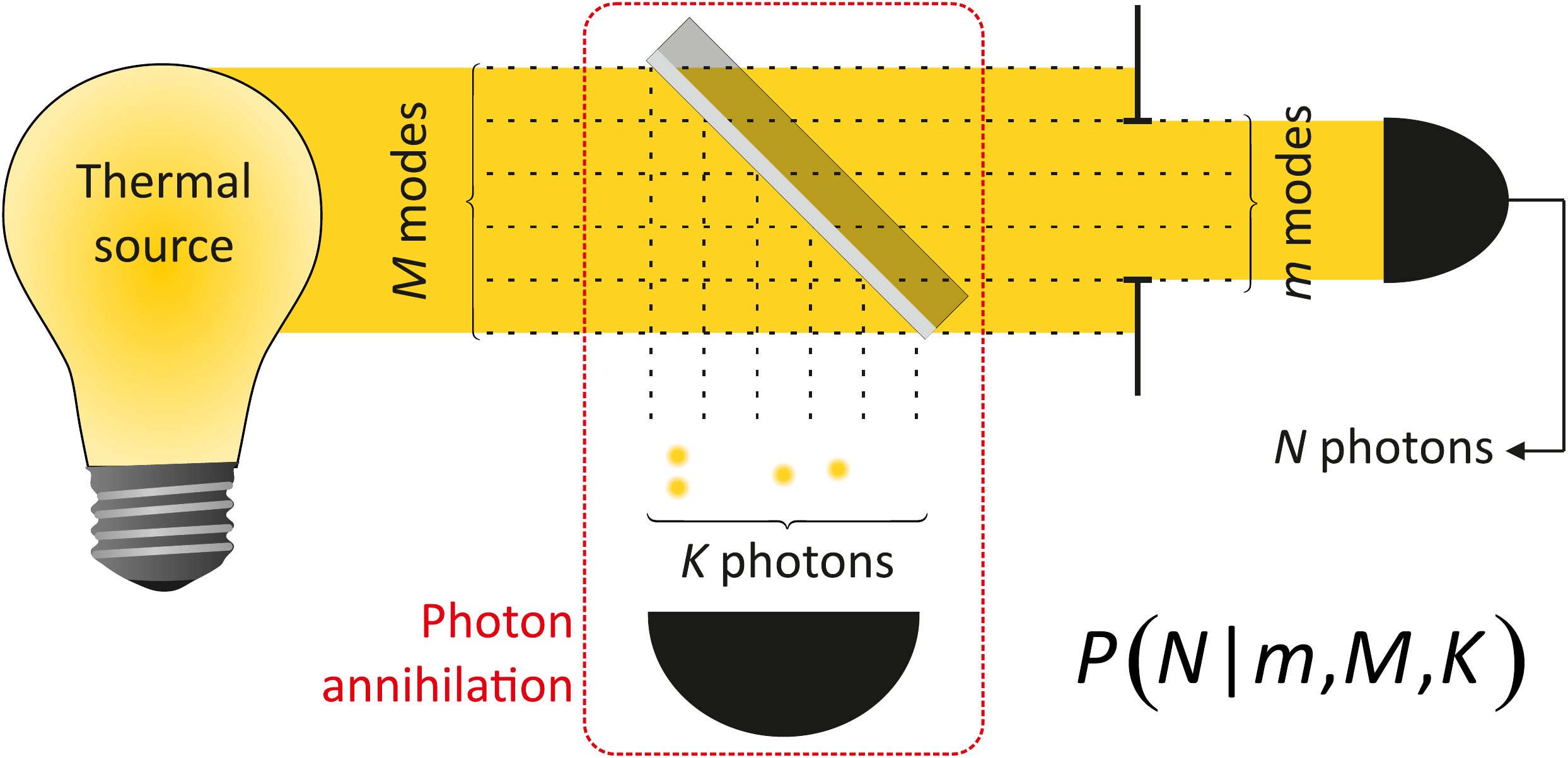}}
		\caption{(Color online). The multiphoton-subtracted multimode thermal state photon number statistics registration scheme.}
		\label{fig:1}
	\end{figure}

	The general case of the multiphoton-subtracted multimode thermal state photon number statistics registration scheme is presented in Fig.\ref{fig:1}. We start from the $ M $-mode thermal state
	\begin{equation}\label{eq:rho_mts}
		\hat{\rho}_{MTS}=\bigotimes\limits_{i=1}^{M}\hat{\rho}_{TS_i}=\bigotimes\limits_{i=1}^{M}\sum\limits_{n_i=0}^{\infty}P_{BE}(n_i|\mu_0)\ket{n_i}\!\!\bra{n_i},
	\end{equation}
	where $ i $ is a mode index. This state passes through a low-reflective beam splitter, combined with a photon detector, which presents the usual way of the conditional implementation of photon annihilation \cite{Ourjoumtsev2006,Neergaard-Nielsen2006}. The $ K $-photon detection leads to the $ K $-photon subtracted state:
	\begin{equation}\label{eq:rho_mkts}
		\hat{\rho}_{MKTS}=\sum\limits_{k_1+\cdots+k_M=K}^{}P(k_1,\ldots,k_M|K)\bigotimes\limits_{i=1}^{M}\hat{\rho}_{k_iTS},
	\end{equation}
	where $ k_i $ denotes the number of subtracted photons in each $ i $ mode, $ P(k_1,\ldots,k_M|K) $ is the probability that exactly $ k_1,k_2,\ldots,k_M $ photons are subtracted in the modes $ 1,2,\ldots,M $ respectively under the condition that the total number of subtracted photons equals $ K $. The first sum is taken over all the indices $ k_i $  combinations, satisfying this condition. Finally, the total photon number $ N $ of the $ m $-mode subsystem of (\ref{eq:rho_mkts}) is measured.

	The paper is organized as follows. In Sec. 1. we derive the photon number distribution for $ m $-mode subsystem of $ K $-photon subtracted $ M $-mode thermal state $ P_N(N|K,M,m,\mu_0) $. In Sec. 2 we describe the experimental verification of this distribution model. In Sec. 3 we present the comparison between the measured data and the theoretical model. Finally, we summarize the obtained results in Conclusion section.

\section{1. Theory}
	The equivalence between mode addition and photon subtraction for thermal states of light can be easily shown with use of probability generating function $ G(z)$,
which is linked to the photon number distribution $P(n)$ as follows:
\begin{equation}\label{eq:G_Pn}
G(z)\equiv\sum\limits_{n=0}^{\inf}P(n)z^n,\quad P(n)=\frac{1}{n!}\left.\frac{\partial^n G(z)}{\partial z^n}\right|_{z=0}
\end{equation}

	 The generating function of the initial thermal state (\ref{eq:rho_ts}) $ P_{BE}(n|\mu_0) $ equals
	\begin{equation}\label{eq:G_BE}
		G_{BE}(z|\mu_0)=\left[1+\mu_0(1-z)\right]^{-1},
	\end{equation}
and the generating function of the compound Poisson distribution (\ref{eq:Pcp}) $ P_{cP}(n|\mu_0,a) $ equals
	\begin{equation}\label{eq:G_cpp}
		G_{cP}(z|\mu_0,a)=\left[1+\mu_0(1-z)\right]^{-a}.
	\end{equation}
		Addition of one more thermal mode leads to the multiplication of the generating function by $ G_{BE}(z|\mu_0) $. Therefore, the $ m $-mode thermal state generating function is
	\begin{equation}\label{eq:G_MTS}
	G_{MTS}(z|\mu_0,m)=\left[1+\mu_0(1-z)\right]^{-m}\equiv G_{cP}(z|\mu_0,m).
	\end{equation}
	
	As was shown in \cite{Bogdanov2017,Bogdanov2016} the photon subtraction can be described by a normalized generating function derivative:
	\begin{equation}\label{eq:G_1}
		G_{-1}(z)=\frac{G^{(1)}(z)}{G^{(1)}(1)}.	
	\end{equation}
	Substituting (\ref*{eq:G_MTS}) into (\ref{eq:G_1}) we obtain $ G_{cP}(z|\mu_0,m+1) $. Thus the $ k $-photon subtracted $ M $-mode thermal state has the function $ G_{cP}(z|\mu_0,m+k) $. Thereby, mathematically the equivalence between mode addition and photon subtraction is linked to the equivalence between exponentiation and normalized differentiation for the generating function (\ref{eq:G_BE}).
	
	Next, let's move on to the general case, where the $ K $ photon subtraction takes place in   $ M\geq m $ modes, and we consider the photon number distribution of its $ m $-mode subsystem $ P_N(N|K,M,m,\mu_0) $. 
	To calculate it, we need to multiply the previous photon number distribution $P_{cP}(N|\mu_0,a=k+m)$ by the probability $P_k(k|K,M,m) $, that exactly $k$ photons are subtracted from the considered subsystem, while the total number of subtracted photons from all the $ M $ modes equals $ K $. Finally, we sum up the probabilities of all outcomes with different $k$ and obtain the following equation:
	\begin{equation}\label{eq:P_N}
		\begin{split}
			P_N(N|K,M,m,\mu_0)=\sum\limits_{k=0}^{K}&P_k(k|K,M,m)\times \\
			\times&P_{cP}(N|\mu_0,a=k+m).
		\end{split}
	\end{equation}
 The probability $P_k(k|K,M,m)$ can be found in terms of balls and boxes combinatorial problem \cite{Landau1958}. The number of ways in which $ K $ identical balls can be distributed into $ M $ distinct boxes ($ K $ photons can be subtracted from $ M $ modes) equals $ C_{M+K-1}^K $. Similarly, the number of ways in which $ k $ photons can be subtracted from $ m $ modes is $ C_{m+k-1}^k $  and for $ K-k $ photons that can be subtracted from $ M-m $  modes the number equals $ C_{M-m+K-k-1}^{K-k} $. Therefore, the required probability
	\begin{equation}\label{eq:P_k}
		P(k|K,M,m)=\frac{C_{m+k-1}^k C_{M-m+K-k-1}^{K-k}}{C_{M+K-1}^K}.
	\end{equation}
	This is a Polya distribution \cite{Bogdanov2003,Avosopiants2019,Johnson1977} which has the generating function \cite{Bogdanov2003}:
	\begin{equation}\label{eq:G_k}
		G(z|K,M,m)={}_2F_1(-K,m,M,1-z),
	\end{equation}
	where $ {}_2F_1 $i s a Gauss hypergeometric function.
	
	The factor $ P_{cP}(N|\mu_0,k+m) $  in (\ref{eq:P_N}) can be interpreted as photon number distribution of multimode thermal state with a random mode number $ k+m $  (the sum of random variables subjected to the distribution (\ref{eq:G_BE}) with a random number of summands), where $ m $ is constant and $ k $ is subject to the Polya distribution (\ref{eq:P_k}). Thus, the generating function of the required distribution (\ref{eq:P_N}) can be obtained by the generating function composition rule as a compound generating function \cite{Feller1968}:
	\begin{equation}\label{eq:G_z}
		\begin{split}
			G(z|K,M,&m,\mu_0)=(G_{BE}(z|\mu_0))^m\times\\
			&\times{}_2F_1(-K,m,M,1-G_{BE}(z|\mu_0)).
		\end{split}
	\end{equation}
	The corresponding photon number distribution can be directly calculated according to (\ref{eq:G_Pn}) and has the following form:
	\begin{equation}\label{eq:P_N_full}
		\begin{split}
			P_N(N|K,M,m,\mu_0)=\frac{\mu_0^N}{(1+\mu_0)^{N+m}}\times\\
			\times\frac{1}{\Gamma(m)}\frac{\Gamma(N+m)}{\Gamma(N+1)}\frac{\Gamma(M)}{\Gamma(M-m)}\frac{\Gamma(M+K-m)}{\Gamma(M+K)}\times\\
			\times{}_2F_1\left(-K,N+m,-K-M+m+1,\frac{1}{1+\mu_0}\right).
		\end{split}
	\end{equation}
	The considered generating function (\ref{eq:G_z}) generates the probability distribution (\ref{eq:P_N_full}) only when $ m<M $.
	
	The considered distribution has the following properties.
	\begin{enumerate}
		\item When $ m \longrightarrow M $ it turns to the compound Poisson distribution (\ref{eq:Pcp}):
			\begin{equation}\label{eq:P_N_M}
				P_N(N|K,M,m=M,\mu_0)=P_{cP}(n|\mu_0,K+M).
			\end{equation}
		\item For $ K=0 $ it also turns to the compound Poisson distribution and does not depend on $ M $:
			\begin{equation}\label{eq:P_N_M_k}
				P_N(N|K=0,M,m,\mu_0)=P_{cP}(N|\mu_0,m).
			\end{equation}
		\item Its mean photon number
			\begin{equation}\label{eq:mu}
				\mu=m\mu_0\left(1+\frac{K}{M}\right).
			\end{equation}
		\item Its correlation function
			\begin{equation}\label{eq:g2}
				g^{(2)}(0)=\frac{1+1/m}{1+1/M}\left(1+\frac{1}{M+K}\right).
			\end{equation}
		\item In contrast to the compound Poisson distribution $ P_{cP}(N|\mu_0,k+1) $, the distribution (\ref{eq:P_N_full})  allows only an integer number of subtracted photons $ K $ (otherwise the distribution diverges).
	\end{enumerate}
\section{2. Experiment}
\begin{figure}[!h]
		\center{\includegraphics[width=0.8\columnwidth]{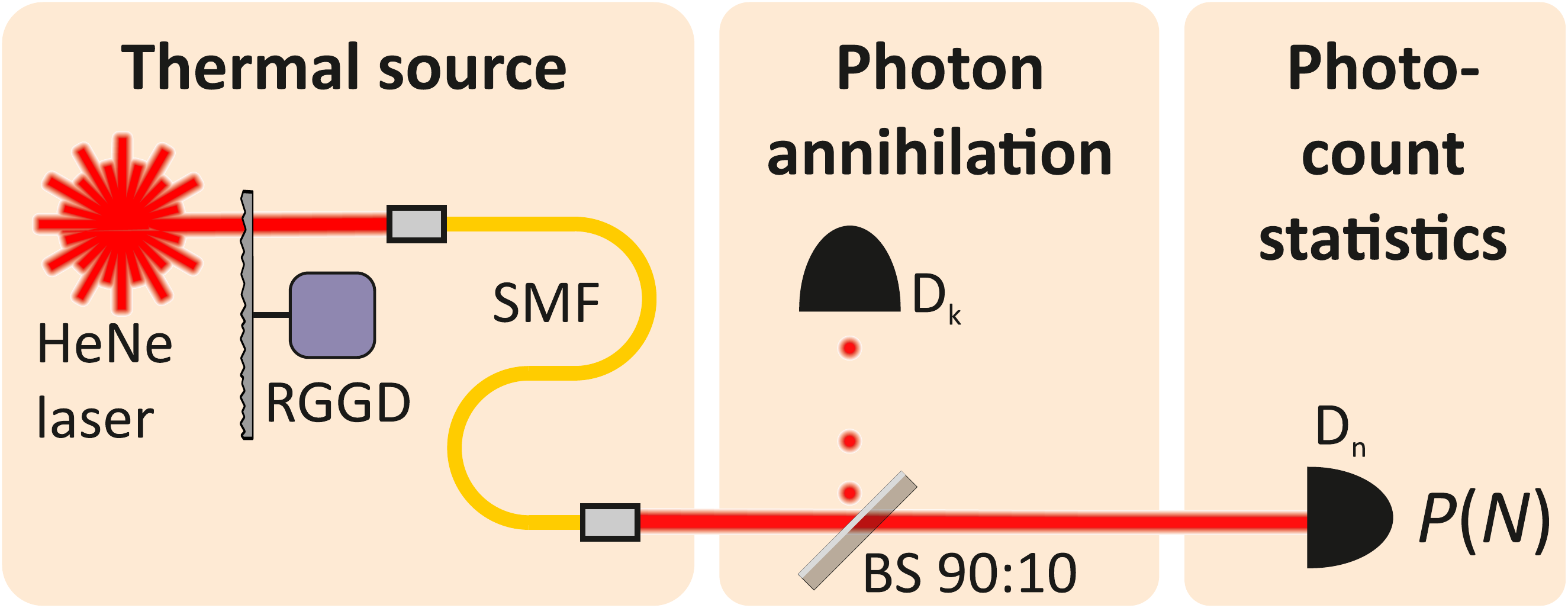}}
		\caption{(Color online). The experimental setup. BS -– beam splitter, RGGD –- rotating ground glass disk, SMF –- single-mode fiber, $ D_k $ and $ D_n $ are single-photon APD-based detectors used for photon annihilation and photo-count statistics measurement respectively.}
		\label{fig:2}
	\end{figure}

	The sketch of our experimental setup is presented in Fig.\ref{fig:2}., which is quite similar to the setup described in \cite{Bogdanov2017}. The HeNe cw radiation was passed through a rotated ground glass disk (RGGD) and transmitted through a single-mode fiber (SMF) for single-mode thermal state preparation \cite{Arecchi1965,Martienssen1964}. A small part of the fiber output beam was redirected by a 90:10 beam splitter to a single-photon detector $ D_K $ based on the silicon avalanche photodiode (APD), in order to implement conditional photon annihilation \cite{Ourjoumtsev2006,Neergaard-Nielsen2006}. Next the multiphoton subtracted thermal light was directed into another single-photon detector $ D_N $ for photo-count distribution $ P(N) $  measurement.
	\begin{figure}[!h]
		\center{\includegraphics[width=\columnwidth]{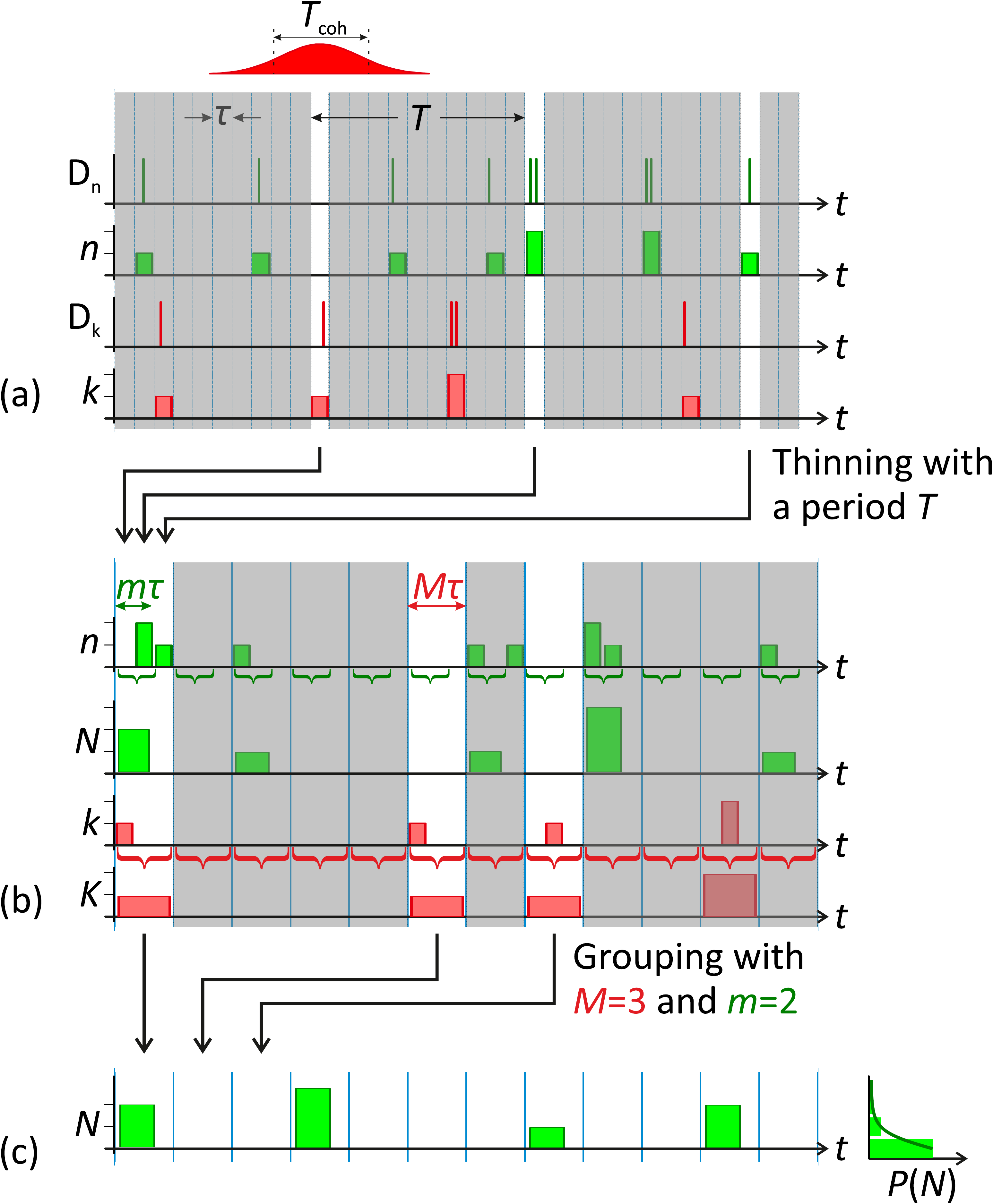}}
		\caption{(Color online). Signal processing. (a) Initial data set is divided by time bins $ \tau $  and then they are thinned with a period $ T $ in order to avoid interbin correlations. (b) Thinned data are grouped by $ M $ and groups are sorted according to the total number of subtracted photons in the group $ K $. The total photon number $ N $ is calculated as a sum of the first $ M\leq m $ bins. (c) Finally the histogram of $ \{N\} $, corresponded to the same value of $ K $  are collected and fitted with the probability distribution $ P_N(N|K,M,m,\mu_0) $ (\ref{eq:P_N_full}.}
		\label{fig:3}
	\end{figure}

	Thus, photo-count pulses from $ D_n $ and $ D_k $ have been synchronously collected. The data processing algorithm is presented in Fig.\ref{fig:3}. First, all the time traces have been divided by time bins with the width $ \tau $  corresponding to the time mode duration (Fig.\ref{fig:3}a). The value of $ \tau $  should satisfy the inequality $ T_{coh}\gg\tau\gg\tau_d $ , where $ T_{coh} $  is the thermal state coherence time defined by the RGGD velocity and $ \tau_d $  is a single-photon detector dead time. This inequality defines the possibility of several photo-counts registration belonging to the same optical mode (see  \cite{Bogdanov2017} for details). In our experiment $ T_{coh}=40~\him{\mu s} $ , $ \tau_d=220~\him{ns} $ and $ \tau=10~\him{\mu s} $,  so the inequality has been satisfied. For each time bin the photo-count numbers $ k $ and $ n $ from the detectors $ D_k $ and $ D_n $ correspondingly have been calculated. Next, in order to avoid any interbin correlations, we selected the bins periodically separated by $ T=12T_{coh} $.

	Since the thermal state was spatially single-mode, the only one way to perform multi-mode state preparation was to collect of $ M $ time modes. Therefore, all the uncorrelated time bins have been grouped by $ M $ Fig.\ref{fig:3}b), and for each group the total number of subtracted photons  $ K $ has been obtained. In order to analyze the situation described in Fig.\ref{fig:1}, where just a part of the thermal modes is finally collected, we calculated the total photon number $ N $ as a sum of the first $ m $ bins in a group.
	
	Considering the $ K $-photon subtracted state we selected the groups with the total number of annihilated photons $ K $ (Fig.\ref{fig:3}c). Thus, for each value of $ M=1\div5 $, $ m=1\div M $ and $ k=0\div5 $  we derived a set of photo-count values $ \{N\} $. The histogram of collected data was compared with the distribution (\ref{eq:P_N_full}).

	However, $ P(N) $  does not exactly equal the measured photo-count distribution because the latter also includes dark counts $ N_D $, described by the Poisson distribution $ P(N_D) $ with the mean value $ \mu_D=m\times 0.0015 $, while the mean photon number per mode $ \mu_0=0.24 $. Thus the distribution model which describes the data $ \{N\} $ is a convolution of (\ref{eq:P_N_full}) and $ P(N_D) $.

\section{3. Results}

	\begin{figure}[!]
		\center{\includegraphics[width=0.9\columnwidth]{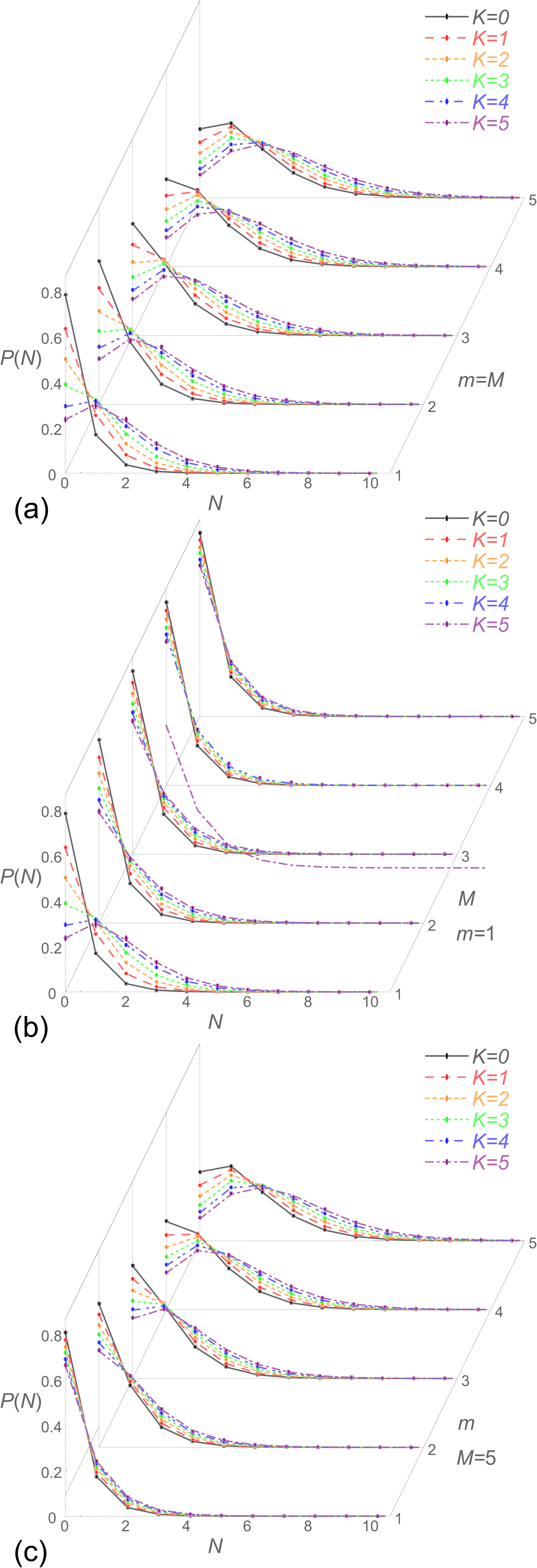}}
		\caption{(Color online). The photon number distribution $ P_N $ (\ref{eq:P_N_full}) for various $ m=M $ (a),  for various $ M $ and fixed $ m=1 $ (b),  for various $ m $ and fixed $ M=5 $ (c). Different number of subtracted photons $K$ is denoted by the line style: from top to bottom at the $y$ $(N=0)$ axis: $K=0,\dots K=5$.}
		\label{fig:4}
	\end{figure}

	Some examples of collected histograms of the photo-count number $ \{N\} $ and corresponding model distributions $ P_N(N|K,M,m,\mu_0) $ are presented in Figure \ref{fig:4}. For all the plots dots with statistical error bars correspond to the measured data and lines – to the model (\ref{eq:P_N_full}) (including dark counts). Different number of subtracted photons $ K $ is denoted by line styles – from top to bottom on the $ y $ $(N=0)$axis:  $K=0,\dots K=5$.

%

	In Fig.\ref{fig:4}~(a) we present the distributions $ P_N $  (\ref{eq:P_N_full}) at $ m=M $. As was shown above, it equals compound Poisson distribution (\ref{eq:P_N_M}), where the mode addition is equivalent to the photon subtraction. In Fig.\ref{fig:4}~(b) we present the single-mode photon number distributions for $ m=1 $ and $ M=1\div5 $. One can see that with increasing $ M $ the action of the photon subtraction become weaker. 

\begin{figure*}[t!]
		\center{\includegraphics[scale=0.3]{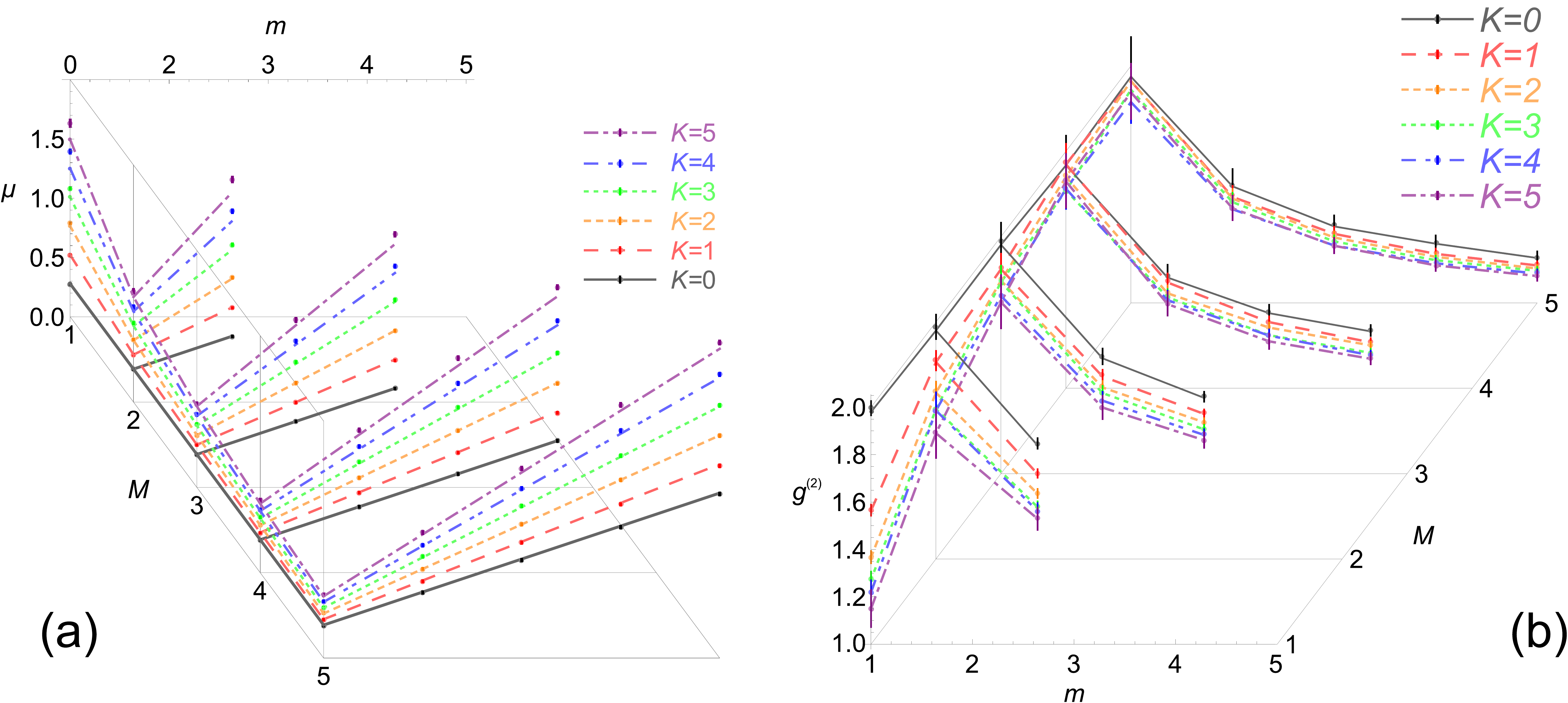}}
		\caption{(Color online). The mean photon number $ \mu $ (a) and the correlation function $ g^{(2)}(0) $ (b) for various total mode number $ M $, observed mode number $ m $ and number of subtracted photons $ K $. Measured points are compared with theoretical lines (\ref{eq:mu}). Different number of subtracted photons $K$ is denoted by the line style: from top to bottom: $K=5,\dots K=0$ in (a) and $K=0,\dots K=5$ in (b).}
		\label{fig:7}
	\end{figure*}

In Fig.\ref{fig:4}~(c) we present the reverse set, where the full mode number $ M $ is fixed and equals 5, and observed mode number $ m $ is increasing from 1 to 5. Again one can note that with the increase of the observed mode fraction $ m/M $ the difference between the states with different photon subtraction become more significant.

	For all the measured histograms our model (\ref{eq:P_N_full}) passed the convenient adequacy $ \chi^2 $  test at the significance level $ p=0.05 $ . The sample size of all states ranged from $ 2000 $ to $ 20000 $.

	In addition we provide the plots of the mean photon number $ \mu $  (Fig.\ref{fig:7}a) and correlation function $ g^{(2)}(0) $ (Fig.\ref{fig:7}b), where one can note a good agreement between the measured data and theoretical predictions (\ref{eq:mu}-\ref{eq:g2}), including the dark counts.

%


\section{Conclusion}
	Thus we have found a photon distribution~(\ref{eq:P_N_full}) based on the combination of compound Poisson~(\ref{eq:Pcp}) and Polya~(\ref{eq:P_k}) distributions, describing the $ m $-mode subsystem of $ M $-mode thermal state under $ K $-photon subtraction (Fig.\ref{fig:1}). We have  studied its some particular cases and derived the simple equations for its mean photon number $ \mu $  (\ref{eq:mu}) and correlation function $ g^{(2)}(0) $ (\ref{eq:g2}). We have analyzed the considered situation experimentally (Fig.\ref{fig:2},~\ref{fig:3}) and obtained the photo-count histograms (Fig.\ref{fig:4}), which are in a good agreement with the presented theoretical model, as well as the measured mean photon number and correlation function (Fig.\ref{fig:7}).

\section{Acknowledgments}
The work is supported by Russian Science Foundation (RSF), project no: 19-72-10069.

Authors express their gratitude for the helpful discussions to Boris Bantysh.

\bibliographystyle{apsrev4-1}   	
\bibliography{MyCollection3}

\end{document}